\documentclass[superscriptaddress,twocolumn,amsmath,amssymb]{revtex4}
\usepackage{graphicx}
\usepackage{dcolumn}
\usepackage{bm}
\usepackage{latexsym}
\begin{document}
\title{Traffic of interacting ribosomes: effects of single-machine 
mechano-chemistry on protein synthesis} 
\author{Aakash Basu}
\affiliation{Department of Physics, Indian Institute of Technology,
Kanpur 208016, India.}
\author{Debashish Chowdhury{\footnote{Corresponding author: debch@iitk.ac.in}}}
\affiliation{Department of Physics, Indian Institute of Technology,
Kanpur 208016, India.}
\date{\today}%

\begin{abstract}
Many {\it ribosomes} simultaneously move on the same messenger RNA 
(mRNA), each separately synthesizing the protein coded by the mRNA. 
Earlier models of ribosome 
traffic represent each ribosome by a ``self-propelled particle'' 
and capture the dynamics by an extension of the totally asymmetric 
simple exclusion process (TASEP). In contrast, here {\it we develope a 
theoretical model} that not only incorporates the 
{\it mutual exclusions} of the interacting ribosomes, but also 
describes explicitly the mechano-chemistry of each of these 
individual cyclic machines during protein synthesis. Using  
analytical and numerical techniques of non-equilibrium 
statistical mechanics, we analyze this model and illustrate its 
power by making experimentally testable predictions on the rate 
of protein synthesis in real time and the density profile of the ribosomes on 
some mRNAs in {\it E-Coli}.  
\end{abstract}.

\maketitle

{\it Translation}, the process of synthesis of proteins by decoding genetic 
information stored in the mRNA, is carried out by {\it ribosomes}. 
Understanding the physical principles underlying the mechanism of operation 
of this complex macromolecular machine \cite{spirin02} will not only provide  
insight into the regulation and control of protein synthesis, but may also 
find therapeutic applications as ribosome is the target of many antibiotics 
\cite{hermann}. 

Most often many ribosomes move simultaneously on the same mRNA 
strand while each synthesises a protein. In all the earlier models of 
collective traffic-like movements of ribosomes 
\cite{macdonald69,lakatos03,shaw03,shaw04a,shaw04b,chou03,chou04}, 
the entire ribosome is modelled as a single ``self-propelled particle'' 
ignoring its molecular composition and architecture. Moreover, in 
these models the inter-ribosome interactions are captured through 
hard-core mutual exclusion and the dynamics of the system is formulated 
in terms of rules that are essentially straightforward extensions of 
the TASEP \cite{schuetz}. 

In reality, the mechanical movement of each ribosome is coupled to 
its biochemical cycle. The earlier TASEP-like models 
cannot account for those aspects of spatio-temporal organization 
that depend on the detailed mechano-chemical cycle of each ribosome. 
In this letter we develope a ``unified'' model that not only 
incorporates the hard-core mutual exclusion of the interacting 
ribosomes, but also captures explicitly the essential steps in the 
biochemical cycle of each ribosome, including GTP (guanine 
triphosphate) hydrolysis, and couples it to its mechanical movement 
during protein synthesis. Consequently, in the low-density limit, 
our model accounts for the protein synthesis by a single isolated 
ribosome while at higher densities the same model predicts not only 
the rate of protein synthesis but also the collective density profile  
of the ribosomes on the mRNA strand. 

\begin{figure}[]
\begin{center}
~~~~~~~~~~~~~~~~~~~~~~~~~~~~~~~~~~~~~~~~~~~~~~~~~~~~~~~~~~~~~~(a)
\includegraphics[width=0.77\columnwidth]{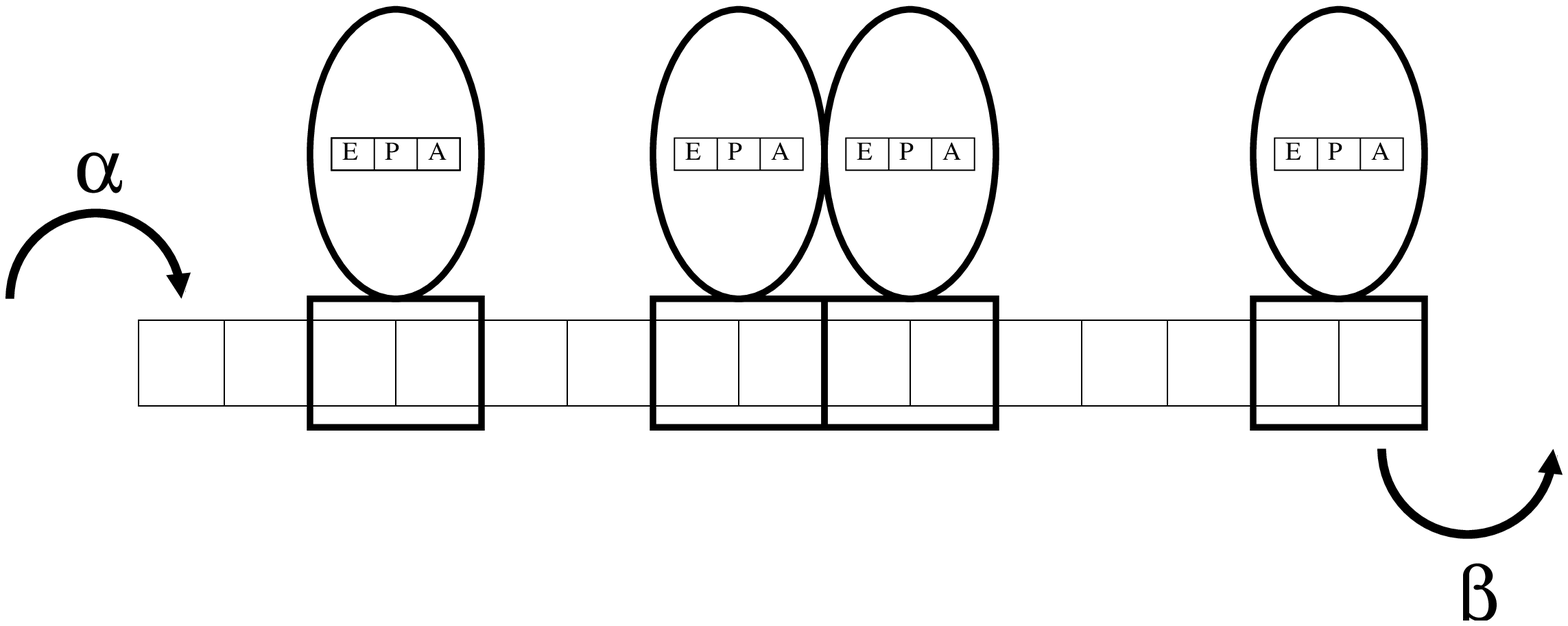}
(b)
\includegraphics[width=0.68\columnwidth]{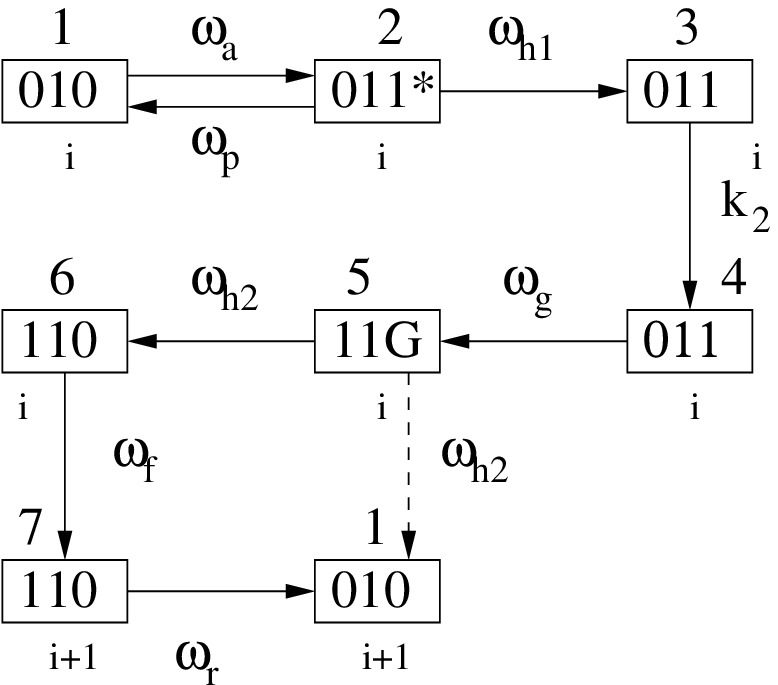}
\end{center}
\caption{(a) The mRNA is represented by a one-dimensional lattice 
each site of which corresponds to a distinct codon; the smaller 
subunit of the ribosome (represented schematically by the rectangle) 
can cover simultaneously ${\ell}$ codons (${\ell} = 2$ in this figure) 
while the letters E, P, A denote the three binding sites on its larger 
subunit. The parameters $\alpha$ and $\beta$ capture the effective 
rates of initiation and termination of translation. (b) The biochemical 
cycle of a single ribosome during the elongation stage. Each box 
represents a distinct state of the ribosome. The index below the box 
labels the codon on the mRNA with which the smaller subunit of the 
ribosome binds. The number above the box labels the biochemical state 
of the ribosome. Within each box, $1 (0)$ represents presence (absence) 
of tRNA on binding sites E, P, A. $1^{*}$ is a EF-Tu bound tRNA and G 
is a EF-G GTPase. The symbols accompanied by the arrows define the rate 
constants for the corresponding transitions. The dashed arrow represents 
the approximate pathway we have considered in our model.}  
\label{fig-model}
\end{figure}

We represent the mRNA chain, consisting of $L$ codons, by a 
one-dimensional lattice of length $L +{\ell}-1$ where each of the 
first $L$ sites from the left represents a single codon (i.e., a 
triplet of nucleotides). We label the sites of the lattice by 
the integer $i$; the sites $i = 1$ and $i = L$ represent the 
start codon and the stop codon, respectively. 

The smaller sub-unit of the ribosome, which is known to bind with 
the mRNA, is represented by an {\it extended particle} of length 
${\ell}$ (in the units of the size of a codon), as shown in 
fig.\ref{fig-model}(a) (${\ell} = 12$ for all results reported here).
\cite{lakatos03,shaw03,shaw04a,shaw04b,chou03,chou04}, 
Thus, the smaller subunit of each ribosome covers ${\ell}$ codons at 
a time (see fig.\ref{fig-model}(a)). According to our convention, 
the {\it position} of such a ribosome on the mRNA strand will be 
given by the position of the lattice site covered by the {\it left} 
edge of its smaller subunit. Each ribosome moves forward by only 
one site in each step as it must translate successive codons one 
by one. The mutual interactions of the ribosomes translocating on 
the same mRNA is taken into account by imposing the constraint of 
mutual exclusion.

The process of translation itself can be divided into three main stages: 
(a) {\it initiation}, (b) {\it elongation}, and (c) {\it termination}. 
Since our model is {\it not} intended to describe initiation and 
termination in detail, we represent initiation and termination by 
the two parameters $\alpha$ and $\beta$, respectively (see 
fig.\ref{fig-model}(a)). If the first ${\ell}$ sites on the mRNA are 
vacant, this group of sites is allowed to be covered by a ribosome, 
from the pool of unbound ribosomes, with probability $\alpha$ 
in the time interval $\Delta t$ (in all our numerical calculations 
we take $\Delta t = 0.001$ s). Similarly, if the rightmost ${\ell}$ 
sites of the mRNA lattice are covered by a ribosome, i.e., the 
ribosome is bound to the $L$-th codon, the ribosome gets detached 
from the mRNA with probability $\beta$ in the time interval 
$\Delta t$. Moreover, since $\alpha$ is the probability of attachment 
in time $\Delta t$, the probability of attachment per unit time 
(which we call $\omega_{\alpha}$) is the solution of the equation 
$\alpha{}=1-e^{-\omega_{\alpha}\times{}\Delta t}$.

To our knowledge, all the earlier models of ribosome traffic on mRNA 
\cite{macdonald69,lakatos03,shaw03,shaw04a,shaw04b,chou03,chou04}, 
describe elongation also by a single parameter, namely, the 
rate $q$ of hopping of a ribosome from one codon to the next. In 
contrast, we model the mechano-chemistry of elongation in detail 
(fig.\ref{fig-model}(b)). In state $1$, the ribosome begins with a tRNA 
bound to the site P. Binding of a fresh tRNA-EF-Tu complex to site A 
causes the transition $1 \rightarrow 2$. The EF-Tu has a GTP part which 
is then hydrolized to GDP, driving the transition $2 \rightarrow 3$. 
Next, the phosphate group, a product of the hydrolysis, leaves resulting 
in the state $4$. This hydrolysis, finally, releases the 
EF-Tu, and the peptide bond formation becomes possible. After this step, 
the tRNAs shift from site P to E and from site A to P; the site A becomes 
occupied by EF-G, in the GTP bound form, resulting in the state $5$. 
Hydrolysis of the GTP to GDP and the release of EF-G drives the 
transition $5 \rightarrow 6$. The transition $6 \rightarrow 7$ is 
accompanied by conformatinal changes that are responsible for pulling 
the mRNA-binding smaller subunit by one step forward. Finally, the 
tRNA on site A is released, resulting in completion of one biochemical
cycle; in the process the ribosome moves forward by one codon (i.e., one
step on the lattice).

However, in setting up the rate equations below, we treat the 
entire transition $5\rightarrow{}6 \rightarrow{}7 \rightarrow{}1$ 
as, effectively, a single step transition from $5$ to $1$, with 
rate constant $\omega_{h2}$. Thus, throughout this paper we work 
with a simplified model where each biochemical cycle during the 
elongation process consists of {\it five} distinct states. 

The modelling strategy adopted here for incorporating the biochemical 
cycle of ribosomes is similar to that followed in the recent work 
\cite{nosc} on single-headed kinesin motors KIF1A. However, the 
implementation of the strategy is more difficult here not only 
because of the higher complexity of composition, architecture and 
mechano-chemical processes of the ribosomal machinery and but also 
because of the {\it heterogeneity} of the mRNA track \cite{nelson}.

Let $P_{\mu}(i)$ be the probability of finding a ribosome at site 
$i$, in the chemical state $\mu$. Then, 
$P(i) = \sum_{\mu=1}^{5} P_{\mu}(i)$,
is the probability of finding a ribosome at site $i$, irrespective 
of its chemical state. Moreover, if a site is {\it not} covered by 
any part of any ribosome, we'll say that the site is occupied by a 
``hole''.  Furthermore, by the symbol $Q(i|j)$ we denote the 
conditional probability that, given a ribosome at site $i$, there 
is a hole at the site $j$. The master equations for the probabilities 
$P_{\mu}(i)$ are given by
\begin{eqnarray} \label{b}
\frac{dP_{1}(i)}{dt}=\omega_{h2}P_{5}(i-1)Q(i-1|i-1+{\ell}) \nonumber \\
+\omega_{p}P_{2}(i)-\omega_{a}P_{1}(i)\\
(i\ne{}1)\nonumber
\end{eqnarray}
\begin{equation} \label{c}
\frac{dP_{2}(i)}{dt}=\omega_{a}P_{1}(i)-(\omega_{p}+\omega_{h1})P_{2}(i)
\end{equation}
\begin{equation} \label{d}
\frac{dP_{3}(i)}{dt}=\omega_{h1}P_{2}(i)-k_{2}P_{3}(i)
\end{equation}
\begin{equation} \label{e}
\frac{dP_{4}(i)}{dt}=k_{2}P_{3}(i)-\omega_{g}P_{4}(i)
\end{equation}
\begin{eqnarray} \label{f}
\frac{dP_{5}(i)}{dt}=\omega_{g}P_{4}(i)-\omega_{h2}P_{5}(i)Q(i|i+{\ell})\\
(i\ne{}L)\nonumber
\end{eqnarray}
However, the equations for $P_{1}(1)$ and $P_{5}(L)$ have the special forms 
\begin{equation} \label{a}
\frac{dP_{1}(1)}{dt}=\omega_{\alpha}\Big(1-\sum_{s=1}^{{\ell}}P(s)\Big)+\omega_{p}P_{2}(1)-\omega_{a}P_{1}(1)
\end{equation}
\begin{equation} \label{g}
\frac{dP_{5}(L)}{dt}=\omega_{g}P_{4}(L)-\beta{}P_{5}(L).
\end{equation}

\begin{figure} 
\begin{center}
\includegraphics[width=0.9\columnwidth]{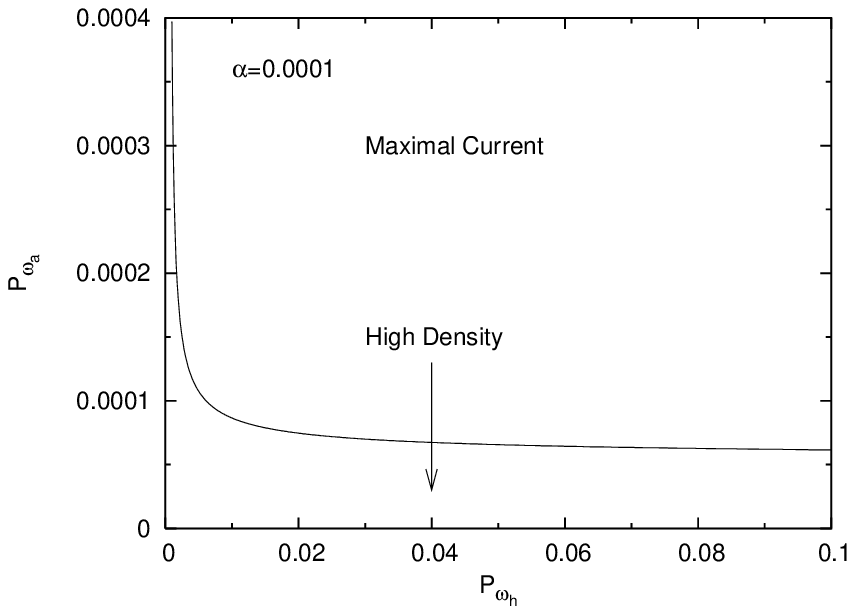}
(a)
\includegraphics[width=0.9\columnwidth]{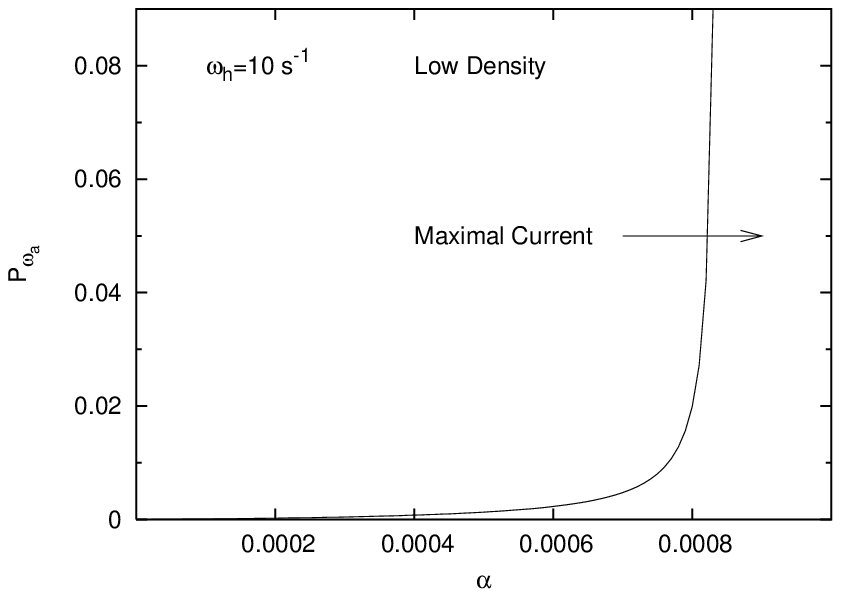}
(b)
\end{center}
\caption{Phase diagram in (a) $P_{\omega_{h}} - P_{\omega_{a}}$ plane 
and (b) $\alpha - P_{\omega_{a}}$ plane. $\omega_g = 25$ s$^{-1}$, 
$\omega_P = 0.0028$ s$^{-1}$, $k_2 = 2.4$ s$^{-1}$, $\beta = 1$.} 
\label{fig7}
\end{figure}

Because of the finite length of the codon sequence between the start 
and stop codons, the open boundary conditions (OBC) are more realistic 
than the periodic boundary conditions (PBC). However, we begin with 
a calculation of the flux of the ribosomes in the steady-state by 
imposing PBC as the results for this artificial situation are required 
for a derivation of the dynamical phase diagram of the system under 
OBC. Under PBC, $P_{\mu}(i)$ for all $i$ are governed by the equations 
(\ref{b})-(\ref{f}). Moreover, under the PBC, only four of the five 
equations (\ref{b})-(\ref{f}) are independent because  
$P(i)=\sum_{\mu=1}^{5}P_{\mu}(i)=N/L = \rho$ 
where $\rho$, the number density of the ribosomes, is a constant 
independent of time; therefore, we do not need to consider  
equation (\ref{b}) for $P_1(i)$ explicitly. In the steady state, 
all time derivatives vanish and because of the translational 
invariance of this state under PBC, the index $i$ can be dropped. 
It is straightforward to show \cite{long} that, for PBC,
\begin{eqnarray} \label{3:2}
Q(i|i+{\ell})= \frac{L-N{\ell}}{L+N-N{\ell}-1}.
\label{eq-qpbc}
\end{eqnarray}
Therefore, under the PBC, equations (\ref{c}-\ref{f}) can be 
solved, using (\ref{eq-qpbc}), to obtain 
\begin{equation}
P_{5} = \frac{P}{1+\frac{\omega_{h2}(L - N{\ell})}{L+N-N{\ell}-1}[\frac{1}{k_{eff}}]}
\end{equation}
where 
\begin{equation}
\frac{1}{k_{eff}} = \frac{1}{\omega_{g}}+\frac{1}{k_{2}}+\frac{1}{\omega_{h1}}+\frac{1}{\omega_{a}}+\frac{\omega_{p}}{\omega_{a}\omega_{h1}}
\end{equation}

The flux of ribosomes $J$, under PBC, obtained from 
$J = \omega_{h2}P_{5}Q(i|i+{\ell})$,
is  
\begin{eqnarray} \label{3:7}
J = \frac{\omega_{h2} \rho (1- \rho {\ell})}{(1+\rho-\rho {\ell})+ \Omega_{h2}(1-\rho {\ell})}
\end{eqnarray}
where $\Omega_{h2} = \omega_{h2}/k_{eff}$. The rate of protein 
synthesis by a single ribosome is ${\ell}J$. This mean-field 
estimate is a reasonably good approximation to the data 
obtained by direct computer simulations \cite{long}. 

It can be shown \cite{long} that, for OBC,  
\begin{eqnarray} \label{Qcalc}
Q(i|i+{\ell}) = \frac{1-\sum_{s=1}^{{\ell}}P(i+s)}{1-\sum_{s=1}^{{\ell}}P(i+s)+P(i+{\ell})}
\end{eqnarray}
and the corresponding flux can be obtained from
\begin{equation}
J=\omega_{\alpha}(1-\sum_{s=1}^{{\ell}}P_{s}) 
\end{equation}

\begin{figure} 
\begin{center}
\includegraphics[width=0.9\columnwidth]{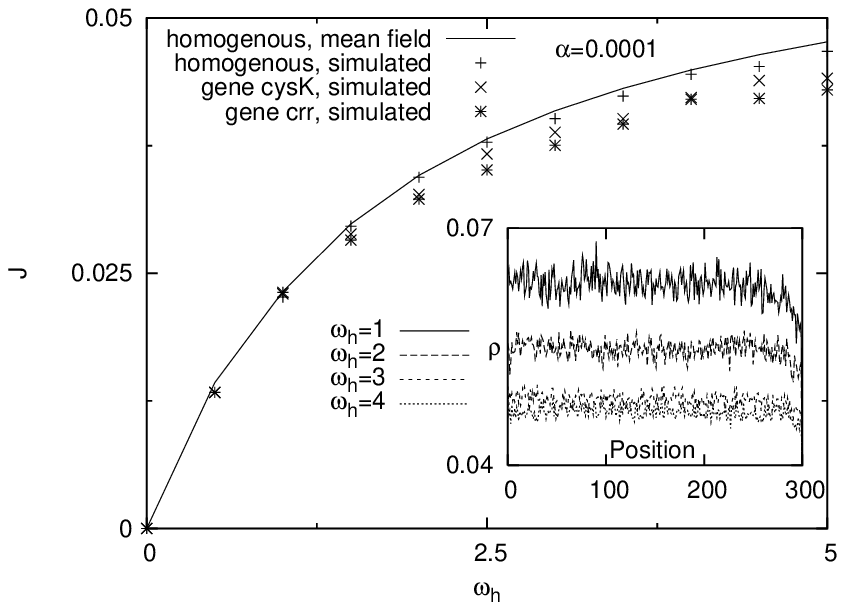}
(a)
\includegraphics[width=0.9\columnwidth]{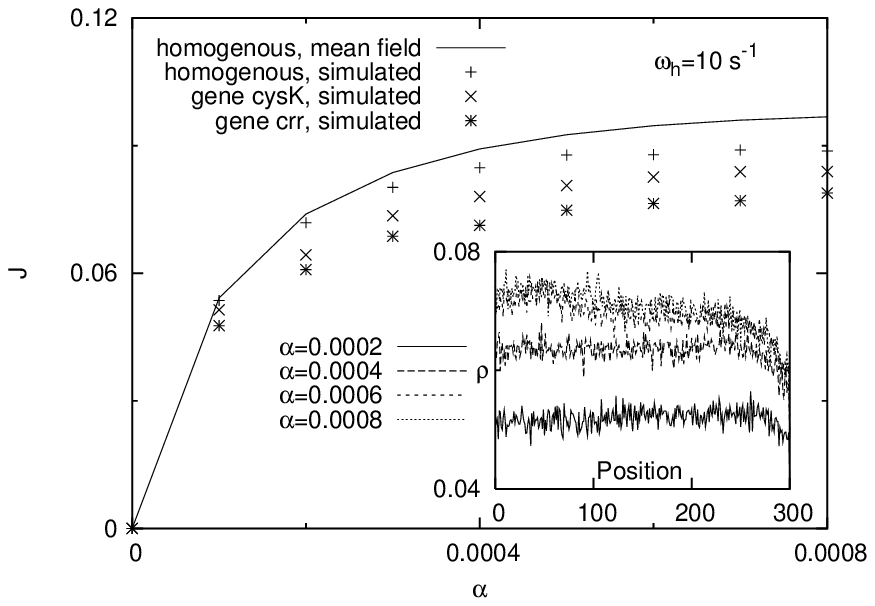}
(b)
\end{center}
\caption{Flux of ribosomes plotted against (a) $\omega_h$ and 
(b) $\alpha$ for the genes  crr (170 codons) and cysK (324 codons) 
of $Escherichia~coli$ K-12 strain MG1655, as well as the 
corresponding curve for a homogenous mRNA strand of 300 codons. 
The insets show the average density profiles on a hypothetical 
{\it homogeneous} mRNA track for four different values of 
(a) $\omega_h$ and (b) $\alpha$, for fixed $\omega_a = 25$ s$^{-1}$.
}    
\label{fig9}
\end{figure}

Motivated by the recent measurements \cite{arava05,mackay04} 
of the number of bound ribosomes on the mRNA, we have computed the 
detailed density profiles of the ribosomes and also drawn the 
phase diagrams in the spirit of the similar plots of non-equilibrium 
dynamical phases of totally asymmetric simple exclusion process 
\cite{schuetz}.

The probabilities $\alpha$ and $\beta$ of initiation and termination 
are incorporated into the model by connecting the ends of the mRNA 
strand to two hypothetical reseroirs with appropriate denities 
$\rho_{-}$ and $\rho_{+}$, respectively \cite{shaw03}. The extremum 
principle \cite{popkov,schuetz} then relates the flux $j$ in the open 
system to the flux $J(\rho)$ for a closed periodic system with the 
same dynamics:
\begin{displaymath}
j = 
\left\{
\begin{array}{ccc}
max~J(\rho) & if~\rho_{-} > \rho > \rho_{+}\\
min~J(\rho) & if~\rho_{-} < \rho < \rho_{+}
\end{array} \right.
\end{displaymath}
For systems with a single maximum in the function $J(\rho)$, 
at $\rho=\rho_{*}$, such as equation (\ref{3:7}), the maximal 
current phase sets in when $\rho_{-} > \rho_{*} > \rho_{+}$. 
By differentiating equation (\ref{3:7}), we find \cite{long}
\begin{equation}
\rho{}_{*}=\frac{-{\ell}\Big(1+\Omega_{h2}\Big)+\sqrt{{\ell}\Big(1+\Omega_{h2}\Big)}}{{\ell}\Big(1-{\ell}-\Omega_{h2}{\ell}\Big)}
\label{soln}
\end{equation}
It can also be shown that \cite{long}
\begin{equation} \label{4:9}
\rho_{-}=\frac{\alpha{}(1-\frac{{\ell}}{L})(1+\Omega_{h2})}{P_{\omega_h} -\alpha{}(1+\Omega_{h2})(1-{\ell})}
\end{equation}
where $P_{\omega_{h}}$ is the probability of hydrolysis in the 
time $\Delta{}t$, and that $\rho_{+}=0$. Similarly, the probability 
of attachment of a $aa-tRNA$ in time $\Delta t$ is denoted by 
$P_{\omega_{a}}$. Thus, the phase boundaries between the various 
phases have been obtained  by solving the equation 
\begin{equation} \label{4:10}
\rho_{-}(\alpha{},\omega_{a},\omega_{h1},\omega_{h2})=\rho_{*}(\alpha{},\omega_{a},\omega_{h1},\omega_{h2})
\end{equation}
numerically, and two typical phase diagrams have been plotted in 
figs.\ref{fig7}(a) and (b) assuming \cite{thompson,harrington} 
$\omega_{h1} = \omega_{h2} = \omega_{h}$.

We focus on genes of $Escherichia~coli$ K-12 strain MG1655 
\cite{databank}. We directly simulate the system by assuming that 
the site dependent transition rate $\omega_{a}$ is proportional 
to the percentage availability of the corresponding aa-tRNA for 
that codon, in the E Coli cell \cite{solomo,andersson90}. In figure 
(\ref{fig9}), we see how the current increases as $\omega_h$ 
(in (a)) and $\alpha$ (in (b)) increases and gradually saturates; 
the saturation value of the current is numerically equal to the 
maximum current obtained in the corresponding case with PBC \cite{long}. 
Simultaneously, the average density of the ribosomes decreases in (a) 
(and increases in (b)) as the parameter $\omega_h$ in (a) (and $\alpha$ 
in (b)) increases, and gradually satuarates. These observations  
are consistent with the scenario of phase transition from one  
dynamical phase to another, as predicted by the extremal current 
hypothesis. Moreover, the lower flux observed for real genes, as 
compared to that for homogeneous mRNA, is caused by the codon 
specificity of the available tRNA molecules.

In this letter we have developed a ``unified'' theoretical model for 
protein synthesis by mutually interacting ribosomes following the 
master equation approach of non-equilibrium statistical mechanics. 
We have computed (i) the rate of protein synthesis in real time and 
(ii) density profile of the ribosomes on a given mRNA, and studied 
their dependences on the rates of various {\it mechano-chemical} 
processes in each ribosome. We have illustrated the use of our model 
by applying these to two genes of {\it E-Coli} and making theoretical 
predictions in real time which, we hope, will motivate new 
{\it quantitative} measurements.

\end{document}